# Shot-noise evidence of fractional quasiparticle creation in a local fractional quantum Hall state

Masayuki Hashisaka[1,*], Tomoaki Ota[1], Koji Muraki[2], Toshimasa Fujisawa[1]

[1]*Department of Physics, Tokyo Institute of Technology, 2-12-1-H81 Ookayama, Meguro, Tokyo 152-8551, Japan*
[2]*NTT Basic Research Laboratories, NTT Corporation, 3-1 Morinosato-Wakamiya, Atsugi, Kanagawa 243-0198, Japan*

We have experimentally identified fractional quasiparticle creation in a tunneling process through a local fractional quantum Hall (FQH) state. The local FQH state is prepared in a low-density region near a quantum point contact (QPC) in an integer quantum Hall (IQH) system. Shot-noise measurements reveal a clear transition from elementary-charge tunneling at low bias to fractional-charge tunneling at high bias. The fractional shot noise is proportional to $T_1(1 − T_1)$ over a wide range of $T_1$, where $T_1$ is the transmission probability of the IQH edge channel. This binomial distribution indicates that fractional quasiparticles emerge from the IQH state to be transmitted through the local FQH state. The study of this tunneling process will enable us to elucidate the dynamics of Laughlin quasiparticles in FQH systems.

Upon the application of a perpendicular magnetic field ($B$), a two dimensional electron system (2DES) forms a fractional quantum Hall (FQH) state at particular rational Landau level filling factors ($\nu = n_e h/eB$; $n_e$: electron density; $h$: Planck's constant; $e$: elementary charge) [1]. Whereas an electron is an elementary particle with charge $e$, an elementary excitation in a FQH state called a Laughlin quasiparticle has a fractional charge $e^* = e/m$, where $m$ is an odd number [2-6]. Fractional quasiparticles have been identified in shot-noise measurements [7-12] performed on a quantum point contact (QPC) in the weak-backscattering regime, where the quasiparticles tunnel through the incompressible FQH liquid [Fig. 1(a)].

When two FQH systems are separated by a vacuum state that acts as a high barrier for fractional quasiparticles, the quasiparticles impinging on the barrier must bunch and rebuild an electron to tunnel [Fig. 1(b)] [13-15]. This bunching process has been experimentally identified in the strong-backscattering regime of a QPC [16,17]. In contrast, we can expect the counterpart process, i.e., creation of fractional quasiparticles, if electrons are forced to tunnel through an incompressible FQH state. Experimentally, this tunneling process can be studied with a local FQH (LFQH) state induced by using a QPC in an integer quantum Hall (IQH) system [Fig. 1(c)]. When the transmission probability is varied by a gate voltage, the increased electrostatic potential simultaneously reduces the local electron density at the QPC. This leads to mismatch between the bulk filling factor $\nu_B$ and the local filling factor $\nu_{QPC}$. In previous reports, the FQH nature of such a LFQH state has manifested itself in the power-law behavior of the tunneling current [18-21]. However, the direct identification of the fractional-charge tunneling has not yet been attained. The complete understanding of the creation of fractional quasiparticles, as well as the bunching of quasiparticles, is highly required to reveal the charge dynamics in quantum Hall systems.

In this study, we identify the creation of fractional quasiparticles in the LFQH state using cross-correlation noise measurements [22]. At high field ($\nu_B \cong 1$), fractional charge ($e/3$) tunneling is observed over a wide range of transmission probability $T_1$ of a $\nu_B = 1$ IQH edge channel. As a function of the bias voltage, the transition of tunneling charge from $e$ to $e/3$ is detected, accompanied by the transition from nonlinear to linear dc transport characteristics. This behavior can be understood from the Tomonaga–Luttinger liquid (TLL) nature of the FQH edge channels. Surprisingly, even when the current is carried by a fully transmitting $\nu_{QPC} = 1/3$ FQH channel, as manifested in the conductance plateau at $e^2/3h$, shot noise is generated indicating the stochastic tunneling of $e/3$ quasiparticles. The reason is that $e/3$ shot noise is generated by the quasiparticle tunneling between $\nu_B = 1$ IQH channels and not between FQH channels, as demonstrated by the $T_1$ dependence of the shot noise.

Measurements were performed at 15 mK in a dilution refrigerator on three QPCs fabricated in an $Al_{0.3}Ga_{0.7}As$/GaAs heterostructure containing a 2DES with an electron density $n_e = 2.3 \times 10^{11}$ cm$^{-2}$ and mobility $\mu = 3.3 \times 10^6$ cm$^2$ V$^{-1}$ s$^{-1}$. In this paper, we show a data set obtained for one of these QPCs in a single cool-down unless otherwise noted. The main results presented in this paper are reproduced in measurements on other QPCs and in different cool-downs. The device has five Ohmic contacts $\Omega_n$ ($n = 1–5$) and a split gate with 200 nm gap. A QPC was formed by applying −0.6 V to one of the split-gate electrodes and varying the other gate voltage $V_g$. A magnetic field was applied in such a direction that the chirality of the edge channels was clockwise, as shown in Fig. 1(d).

Longitudinal $R_{xx}$ and Hall $R_{xy}$ resistance traces of the 2DES, obtained separately, are shown in Fig. 1(e). The small depression features near 5.4 and 6.8 T indicate the incipient formation of the $\nu_B = 5/3$ and $4/3$ FQH states.

We applied a dc voltage $V_1$ to $\Omega_1$ to inject current $I_1$ that is partitioned at the QPC. The reflected (transmitted) current flows to $\Omega_3$ ($\Omega_5$), where only finite frequency (> 1 kHz) noise $\Delta I_3$ ($\Delta I_5$) is collected because of the coupling capacitors placed at the input of the transimpedance amplifiers (TAs) [22]. The dc components $I_2$ and $I_4$ are collected at $\Omega_2$ and $\Omega_4$ located downstream of $\Omega_3$ and $\Omega_5$, respectively. We measured $I_1$ and $I_2$ to evaluate the conductance $G$ of the QPC as $G = I_4/V_1 = (I_1 − I_2)/V_1$. In addition, the differential conductance $g$ was determined as $g =$

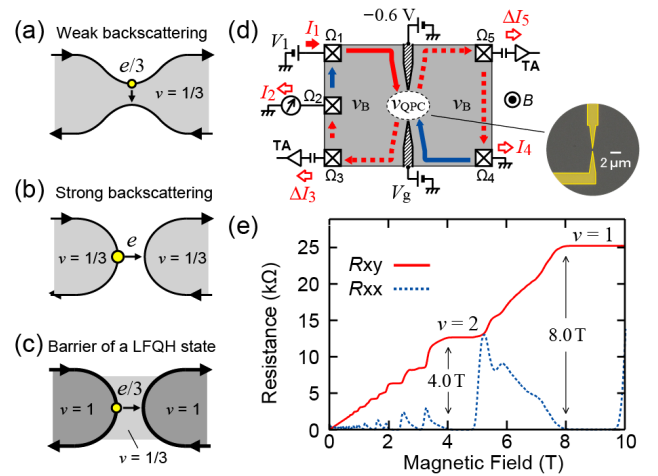

FIG. 1 (color online). (a) Tunneling of $e/3$ quasiparticles in the weak-backscattering regime in a $\nu = 1/3$ FQH state. (b) Electron tunneling in the strong-backscattering regime. (c) $e/3$-charge tunneling through a LFQH state. (d) Schematic of the device and measurement setups. (Inset) Colored optical micrograph of the split gate. (e) Magnetic field dependence of $R_{xx}$ and $R_{xy}$.

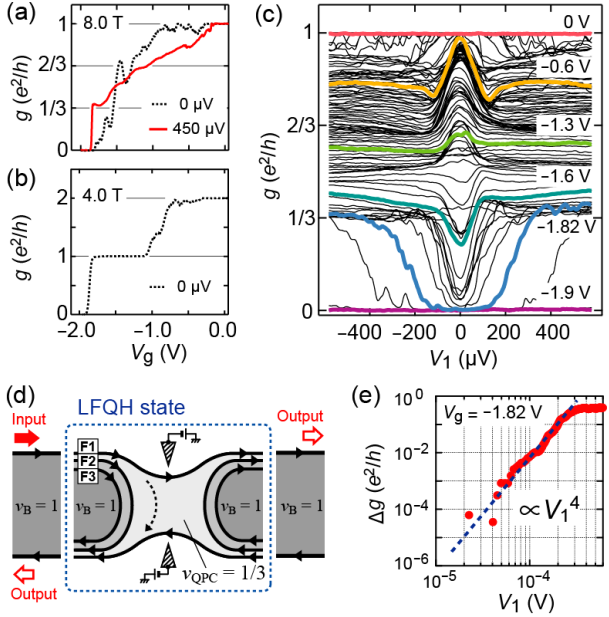

FIG. 2 (color online). (a) $V_g$ dependence of $g$ at $B = 8.0$ T, $V_1 = 0$, and 450 μV. (b) $V_g$ dependence of $g$ at 4.0 T and $V_1 = 0$ μV. (c) $V_1$ dependence of $g$ measured at 8.0 T in 20 mV steps for $V_g$ from 0 V to −2.0 V. Colored thick lines indicate some of the typical traces. (d) Schematic of the edge states near the QPC. F$i$ ($i$ = 1, 2, and 3) labels three FQH channels. (e) Log–log plot of $\Delta g(V_1) = g(V_1) − g(0)$ for the data at $V_g = −1.82$ V.

$dI_4/dV_1 = dI_1/dV_1 − dI_2/dV_1$ using a standard lock-in technique with a small ac modulation (15 μV) of $V_1$ at 19 Hz. We define the transmission probability through the narrow constriction using the relation $G = \Sigma_n T_n \times e^2/h$. Here, $T_n$ is the transmission probability of the $n$th channel. Further details of the measurement setup and analysis are described in Ref. [22].

We verify the LFQH state at high field (8.0 T) in the $V_g$ dependence of $g$ [Fig. 2(a)]. At $V_1 = 0$ μV, $g$ decreases from $e^2/h$ to zero exhibiting many features, which originate from the formation of LFQH states and the resultant TLL nature of the FQH channels [23]. This observation contrasts with the result at 4.0 T, where $g$ decreases smoothly with no additional features below $g = e^2/h$ [Fig. 2(b)], indicating a smooth variation of the barrier potential. At a finite voltage ($V_1 = 450$ μV) where TLL-induced nonlinear behavior disappears, a plateau-like structure is observed at $g \cong e^2/3h$, implying the formation of $\nu_{QPC} \cong 1/3$ LFQH state (for more details, see Supplemental Material [24]).

FQH channels are known to form even in an IQH system because of the gradual electron-density decrease at the edge of the 2DES [28]. These FQH channels copropagate to form an IQH channel along the periphery of the bulk IQH region, and at a QPC they can be separated [Fig. 2(d)] [23,29]. Figure 2(c) shows the $V_1$ dependence of $g$ for different $V_g$. The $g \cong e^2/3h$ plateau appears as accumulated traces at $|V_1| > 200$ μV, indicating the complete transmission of the outermost $\nu_{QPC} = 1/3$ channels [F1 in Fig. 2(d)]. In the low-bias region, the nonlinear behavior of $g$ reflects the suppressed transmission of fractional quasiparticles [arrowed dotted line in Fig. 2 (d)] induced by the TLL nature of the FQH channels. Note that at $V_g = −1.82$ V, $\Delta g = g(V_1) − g(0)$ shows a clear power-law dependence ($\Delta g \propto V_1^\alpha$) with exponent $\alpha = (2/\nu_{QPC}) − 2 = 4$ [Fig. 2(e)], in agreement with the TLL theory for $\nu_{QPC} = 1/3$ [18-21]. These characteristics are a clear indication of the LFQH states.

We evaluated the noise generated at the QPC by measuring the cross correlation $S_{35} = \langle \Delta I_3 \Delta I_5 \rangle$. In the present setup, $S_{35}$ should equal the shot noise $S_{35}^{shot}$, which is theoretically given as follow: [7,22]

$$S_{35}^{shot} = -2e^* I_1 F \left[ \coth\left(\frac{e^* V_1}{2k_B T_e}\right) - \frac{2k_B T_e}{e^* V_1} \right]. \quad (1)$$

In eq. (1), $F = [\Sigma_n T_n(1 − T_n)]/N$ is the shot-noise reduction factor, $k_B$ is the Boltzmann constant, and $T_e$ is the electron temperature. $N$ is the number of channels involved. For a $\nu_B \cong 2$ IQH system, for example, we have $F = [\Sigma_\sigma T_\sigma (1 − T_\sigma)]/2$, where $\sigma$ (= ↑ or ↓) denotes the spin direction.

First, we show $S_{35}$ measured at 4.0 T, where the dc transport shows no signature of a LFQH state. The QPC was set at $g = e^2/3h$ ($V_g = −1.86$ V), where partitioning occurs only in the outer up-spin channel ($T_\uparrow = 1/3$ and $T_\downarrow = 0$). The measured $S_{35}$ is plotted in the inset of Fig. 3(b). With increasing $|V_1|$, $S_{35}$ decreases from zero owing to the generation of shot noise. The data agree well with the $S_{35}^{shot}$ calculated using eq. (1) with $e^* = e$ and $T_e = 82$ mK. This result indicates that at 4.0 T, the QPC works as an ideal beam splitter for incident (up-spin) electrons.

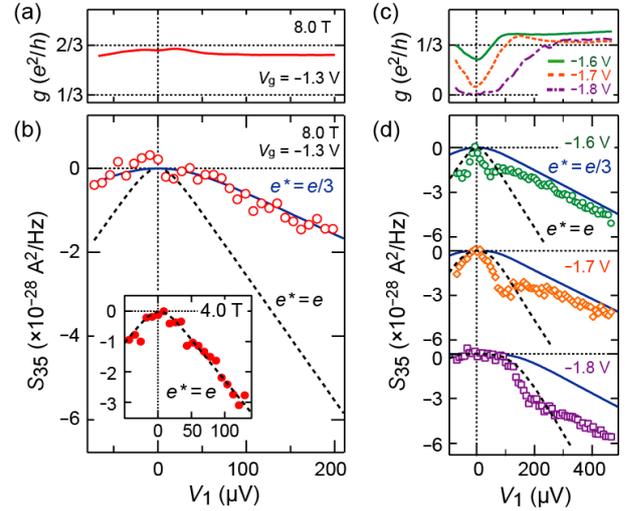

FIG. 3 (color online). (a)(b) $V_1$ dependence of (a) $g$ and (b) $S_{35}$ measured at $B = 8.0$ T and $V_g = −1.30$ V. Solid blue and dotted black lines in (b) are $S_{35}^{shot}$ calculated with $e^* = e/3$ and $e^* = e$, respectively. Inset in (b) shows $S_{35}$ measured at $B = 4.0$ T and $V_g = −1.86$ V, plotted with the trace of $S_{35}^{shot}$ with $e^* = e$. (c) $V_1$ dependence of $g$ measured at $V_g = −1.6, −1.7$, and $−1.8$ V at 8.0 T. (d) $S_{35}$ measured at these values of $V_g$.

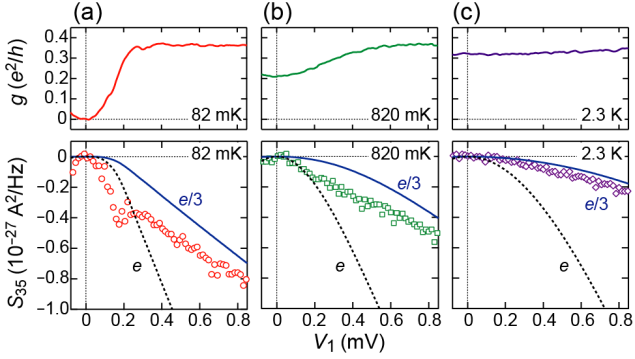

FIG. 4 (color online). $V_1$ dependence of $g$ (upper) and $S_{35}$ (lower panels) measured at $B = 8.8$ T and (a) $T_e = 82$ mK, (b) 820 mK, and (c) 2.3 K. These data were taken from the same sample in a different cooldown, with $-0.7$ and $-1.6$ V applied to the split-gate electrodes to form a QPC with $T_1 \cong 1/3$.

At 8.0 T, we observed markedly different behavior reflecting the formation of a LFQH state. Figures 3(a) and 3(b) show $g$ and $S_{35}$, respectively, as a function of $V_1$. At this gate voltage ($V_g = -1.30$ V), $g$ remains almost constant over the entire $V_1$ range. $S_{35}$ decreases with increasing $|V_1|$ in a similar manner as that at 4.0 T; however, the data are better fitted by eq. (1) with $e^* = e/3$ rather than $e^* = e$. Note that we consider $F$ in eq. (1) as $F = T_1(1 - T_1)$.

At gate voltages where $g$ exhibits power-law $V_1$ dependence, a transition of the tunneling charge from $e$ to $e/3$ was observed. Figures 3(c) and 3(d) show $g$ and $S_{35}$, respectively, measured at $V_g = -1.6$, $-1.7$, and $-1.8$ V. The behavior of $S_{35}$ at low bias is well described by eq. (1) with $e^* = e$, whereas $S_{35}$ departs from the curve at high bias, exhibiting a much weaker $V_1$ dependence that is again in accordance with $e^* = e/3$. Here again, we consider $F = T_1(1 - T_1)$. Notably, for each $V_g$, the transition between $e^* = e$ and $e/3$ occurs at the same $V_1$ as the onset of the power-law suppression of $g$. This transition is observed not only at these $V_g$ values but also over a wide range of $V_g$ (Supplemental Material [24]); the shot noise of $e/3$ quasiparticles is generally measured in the linear conductance regime at high bias, while $e^* = e$ is obtained in the nonlinear conductance regime at low bias.

We here note that the $e^* = e$ shot noise in the nonlinear regime is understood from the TLL nature of the FQH channels, which enhances the backscattering of $e/3$ quasiparticles; namely, it suppresses the $e/3$-charge transport through the QPC [see Fig. 2(d)]. In this case, current through the QPC is carried by electron tunneling, which leads to the generation of $e^* = e$ shot noise [30-32]. This explanation is confirmed in the temperature dependence of $g$ and $S_{35}$ (Fig. 4). At high temperatures, where the TLL-induced nonlinear behavior of $g$ disappears, the $e^* = e$ shot noise at low bias is suppressed and $S_{35}$ follows $S_{35}^{shot}$ with $e^* = e/3$ over the entire $V_1$ range (for more details, see Supplemental Material [24]).

We begin the discussion for fractional shot noise in the linear conductance regime at $\nu_{QPC} \cong 1/3$. The formation of FQH channels was confirmed by the $e^2/3h$ plateau in the dc measurements. On the other hand, shot-noise measurements

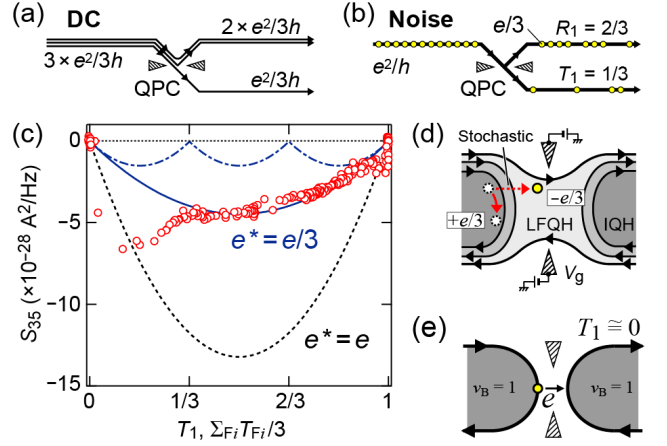

FIG. 5 (color online). (a) Simple schematic of the FQH channels at $\nu_{QPC} \cong 1/3$ expected from the dc transport characteristics. (b) Simple schematic of partitioning $e/3$-quasiparticles expected from the shot noise. (c) $T_1$ dependence of $S_{35}$ measured at $V_1 = 450$ μV and 8.0 T. Solid blue and dotted black lines show $S_{35}^{shot}$ calculated with $e^* = e/3$ and $e$, respectively, considering $F = T_1(1 - T_1)$. Blue dash-dotted line is $S_{35}^{shot}$ with $e^* = e/3$ considering $F = [\Sigma_{Fi} T_{Fi}(1 - T_{Fi})]/3$. (d) Model schematic of the tunneling process. Fractional charges are stochastically generated at the boundary of IQH and LFQH regions, and deterministically fed to the partitioned channels. (e) Electron tunneling between the IQH channels at $T_1 \cong 0$.

indicated $e/3$ quasiparticles. These two experiments may appear contradictory because the $g \cong e^2/3h$ plateau suggests the absence of stochastic process in the FQH channels [Fig. 5(a)], whereas the observed shot noise indicates that the stochastic tunneling of fractional quasiparticles must be occurring [Fig. 5(b)]. The key observation to solve this question is that the shot noise is proportional to $F = T_1(1 - T_1)$. This is demonstrated by plotting $S_{35}$ measured at different $V_g$ as a function of $T_1$ [Fig. 5(c)]. Here, we plotted the data acquired at $V_1 = 450$ μV to focus on the behavior at high bias.

In Fig. 5(c), $S_{35}$ agrees well with eq. (1) calculated for $e/3$-charge. What is important to note is that its variation with $V_g$ is governed by the binomial distribution $-T_1(1 - T_1)$ over a wide $T_1$ range (from 1/3 to 0.9) [blue solid line]. To make this point clearer, we plot the expected behavior if the shot-noise generation is governed by the transmission probability of each fractional channel (blue dash-dotted line), which is given by considering $F = [\Sigma_{Fi} T_{Fi}(1 - T_{Fi})]/3$, where F$i$ ($i = 1$, 2, and 3) labels the three FQH channels. If $T_{Fi}$ varies individually in succession from 1 to 0, $S_{35}$ oscillates as a function of the total transmission probability $\Sigma_{Fi} T_{Fi}/3$ ($= T_1$), as it is expected for partitioning of the three copropagating IQH channels [7]. Obviously, the experimental result differs from the calculation with this assumption. The shot noise governed by the binomial distribution of $T_1$ (and not $T_{Fi}$) indicates that stochastic tunneling occurs in the IQH channels [Fig. 5(b)] and not in the FQH channels. This demonstrates that fractional quasiparticles emerge from the IQH system at the LFQH state. This process arises from the fact that the bulk and constricted regions have different filling factors and thus possess distinct eigenstates

When $T_1$ is close to 0 or 1, $S_{35}$ approaches $S_{35}^{shot}$ for $e^* = e$ [Fig. 5(c)]. This is because, when a large negative $V_g$ is applied to achieve $T_1 \cong 0$, an opaque barrier is established between the two IQH channels, which forbids the transmission of $e/3$ quasiparticles [Fig. 5(e)]. Similarly, at $T_1 \cong 1$, an incompressible IQH state ($\nu_{QPC} \cong 1$) forms between the transmitting IQH channels, inhibiting the reflection of $e/3$ quasiparticles. Thus, at $T_1 \cong 0$ or 1, the tunneling current is carried in units of $e$.

Let us consider the details of the $e/3$-charge tunneling. If the LFQH state acts as a tunnel barrier, the width of the barrier and hence $T_1$ should vary as functions of $V_g$. However, we observed a $g \cong e^2/3h$ plateau, which indicates that $T_1$ is not affected by changes in $V_g$. Although the full mechanism is not yet clear, the following scenario is possible. Near the QPC, the IQH channel splits into three FQH channels, being accompanied by the one-by-one fractional quasiparticle tunneling from the biased IQH region to the LFQH state. This tunneling creates $-e/3$ and $+e/3$ charges, which are composed of the deformation of the LFQH edges [18]. Each of them is deterministically transmitted or backscattered at the QPC [Fig. 5(d)], and therefore $T_1$ does not depend on the width of the LFQH state. Away from the QPC, the three FQH channels again merge into $\nu_B = 1$ IQH channels, and the fractional excitations propagate as edge-magnetoplasmon pulses with $e/3$ charge [33,34]. In this model, shot noise is generated because of the stochastic tunneling at the boundary of IQH and FQH regions. We expect that further investigations will validate this scenario and completely explain the mechanism.

In summary, we experimentally identified the fractional charge tunneling through a LFQH state in a $\nu_B = 1$ IQH system. In the dc transport measurements, the formation of the LFQH state was confirmed in the $g \cong e^2/3h$ plateau, as well as the power-law behavior of the tunneling current. The shot-noise measurements demonstrated the $e/3$-charge tunneling through the LFQH state. The binominal distribution factor $-T_1(1 - T_1)$ indicated that fractional quasiparticles emerge from the IQH system. This quasiparticle creation is regarded as the counterpart of the bunching of quasiparticles, which has been observed in the strong backscattering regime of a QPC in FQH systems.

We appreciate fruitful discussions with N. Kumada, E. Iyoda, T. Kato, D. C. Glattli, S. Ludwig, and T. Martin. We also appreciate the experimental support given by M. Ueki. This study was supported by the Grants-in-Aid for Scientific Research (21000004, 25800176, and 26103508).

---